\documentclass[12pt,preprint]{aastex}

\shorttitle{Magnetic Braking \& Protostellar Disks}
\shortauthors{Mellon \& Li}

\def\lsim{\raise 0.3ex\hbox{$<$}\kern -0.75em{\lower 0.65ex\hbox{$\sim$}}}
\begin{document}

\title{Magnetic Braking and Protostellar Disk Formation: \\  
Ambipolar Diffusion}

\author{Richard R. Mellon$\!$\altaffilmark{1} \&  
Zhi-Yun Li$\!$\altaffilmark{1}}
\altaffiltext{1}{Astronomy Department, P.O. Box 400325, University of Virginia, 
Charlottesville, VA 22904; rrm8p, zl4h@virginia.edu}

\begin{abstract}

It is established that the formation of rotationally supported disks
during the main accretion phase of star formation is suppressed by a
moderately strong magnetic field in the ideal MHD limit. Non-ideal
MHD effects are expected to weaken the magnetic braking, perhaps
allowing the disk to reappear. We concentrate on one such effect,
ambipolar diffusion, which enables the field lines to slip relative
to the bulk neutral matter. We find that the slippage does not
sufficiently weaken the braking to allow rotationally supported disks to
form for realistic levels of cloud magnetization and cosmic ray
ionization rate; in some cases, the magnetic braking is even enhanced.
Only in dense cores with both exceptionally weak fields and
unreasonably low ionization rate do such disks start to form in
our simulations. We conclude that additional processes, such as
Ohmic dissipation or Hall effect, are needed to enable disk
formation. Alternatively, the disk may form at late times when
the massive envelope that anchors the magnetic brake is dissipated,
perhaps by a protostellar wind.

\end{abstract}

\keywords{accretion disks --- ISM: molecular clouds and magnetic 
fields --- MHD --- stars: formation}

\section{Introduction}

%
%

Disk formation is an integral part of star formation that has been 
studied for a long time. Early works concentrated on the collapse
of rotating, non-magnetic cores \citep[e.g.,][]{1984ApJ...286..529T}.  
\citet{1995ARA&A..33..199B} reviewed these works, and noted a number of 
unresolved problems \citep[see also][]{1998ASPC..148..314B}. 
Topping the list is the effect of magnetic
braking on disk formation. 

The magnetic properties of star-forming dense cores are reasonably
well constrained. There is now ample evidence for ordered magnetic
fields on the core scale from polarization of dust emission 
\citep[e.g.,][]{2000ApJ...537L.135W}. A spectacular recent example is the
Submillimeter Array (SMA) observation of NGC 1333 IRS 4A, which shows a pinched magnetic
configuration on the $10^3$~AU scale \citep{2006Sci...313..812G}. The 
magnitude of the magnetic field is harder to determine. 
\citet{2008ApJ...680..457T} carried out an extensive Zeeman survey of 
dark cloud cores in OH, and determined a mean mass-to-flux ratio
of $\lambda \sim 4.8$ in units of the critical value 
$(2\pi G^{1/2})^{-1}$ \citep{1978PASJ...30..671N}; correction for
uncertain projection effects may bring this value to $\sim 2$. 
The implication is that dense cores are typically moderately 
strongly magnetized, with a dimensionless mass-to-flux ratio of
a few to several. A relatively low value of $\lambda \sim 2-3$ 
is expected for dense cores formed out of magnetically subcritical
clouds (with $\lambda < 1$) through ambipolar diffusion 
\citep[e.g.,][]{1989ApJ...342..834L,1994ApJ...432..720B,2008arXiv0804.4201N,2007ApJ...671..497A}.
The values of $\lambda$ are expected to have a wider spread 
for cores formed out of turbulent compression. Nevertheless, 
the cores tend to be more magnetized relative to their masses
than the cloud as a whole, because only a fraction of the mass
along a flux tube that threads the cloud is compressed into the 
core \citep[e.g.,][]{2007ApJ...661..262D,2005prpl.conf.8473T}. It is
unlikely for the cores to have a value of $\lambda$ more than several, 
unless the cloud as a whole is magnetized to an unrealistically 
low level.  
  
%
%
There have been a number of recent calculations that included 
magnetic fields in the collapse of rotating cores, focusing 
mostly on the early phase before the mass of the central object 
reaches the stellar range 
\citep[e.g.,][]{1998ApJ...502L.163T,2006ApJ...647L.151M,2006ApJ...641..949B,2007Ap&SS.311...75P}. 
Other studies have concentrated on the angular momentum evolution and disk formation in
the main accretion phase (\citealp{2003ApJ...599..363A,2006ApJ...647..374G};
\citealp[hereafter Paper I]{2008ApJ...681.1356M}; \citealp{2008A&A...477....9H}). These studies find that
magnetic braking efficiently removes angular momentum from the
collapsing matter, preventing a large, 10$^2$-AU scale rotationally supported
disk from forming in realistic cores in the ideal MHD limit.
However, such disks are routinely observed around young stars, at
least in relatively late, Class I and II phases. How can the disk be saved?

Molecular cloud cores are lightly ionized, so perfect coupling
between the magnetic field and matter assumed in ideal MHD is not
expected. As the density increases, the coupling is weakened first by
ambipolar diffusion, then the Hall Effect, then Ohmic dissipation. In
this paper we will concentrate on ambipolar diffusion as a first
step. The effect of ambipolar diffusion on magnetic braking in
the formation \citep{1994ApJ...432..720B}
and collapse \citep{2002ApJ...580..987K} of rotating cores has been 
examined previously 
using the thin disk approximation. \citet{2002ApJ...580..987K} 
parametrized the braking strength and 
found that disk formation can be suppressed in the ambipolar diffusion 
limit when the braking parameter is large. We use two-dimensional models
which allow direct calculation of the strength of the magnetic
braking. We find that ambipolar
diffusion does not weaken the magnetic braking enough to allow
rotationally supported disks to form under realistic conditions. The
implication is that additional processes must be found to save the
disk; two possibilities are discussed in section 5.

%
%

\section{Model Formulation}

%
%
We study disk formation and evolution during the main accretion phase 
after a central object has formed. The problem setup is identical to 
that of Paper I (to which we refer the reader for details), except 
for the inclusion of ambipolar diffusion. As in \citet{2003ApJ...599..363A}, 
we idealize the initial configuration of the main accretion 
phase as a rotating self-similar singular isothermal toroid 
supported against self-gravity partially by thermal pressure 
and partially by magnetic fields (see figure~1 of Paper I for an 
example). The toroid is characterized by three parameters, the 
isothermal sound speed ($c_s$), mass-to-flux ratio ($\lambda$) and 
rotational speed ($v_0$), which is taken to be spatially constant 
to preserve the self-similarity of the configuration. 

Our treatment of ambipolar diffusion follows that of Shu (1991), where 
the magnetic field is assumed to be tied to ions, which drift relative 
to neutrals with a velocity
\begin{equation}
{\bf v}_d={\bf v}_i-{\bf v}_n=\frac{1}{4\pi\gamma\rho_n\rho_i}
(\nabla\times {\bf B}) \times {\bf B}
\label{drift}
\end{equation}
where $\gamma=3.5\times10^{13}$~cm$^3$~g$^{-1}$~s$^{-1}$ is the ion-neutral drag
coefficient and $\rho_n$ ($\rho_i$) and $v_n$ ($v_i$) are
the density and velocity of the neutrals (ions), respectively. 
We adopt an ion density of the form
\begin{equation}
\rho_i=C~\rho_n^{1/2}
\label{iondensity}
\end{equation}
with the coefficient
\begin{equation}
C=3\times10^{-16}\left(\frac{\zeta}{10^{-17}~s^{-1}}
\right)^{1/2}cm^{-3/2}~g^{1/2}
\label{driftC}
\end{equation}
where $\zeta$ is the cosmic ray ionization rate
\citep{1979ApJ...232..729E}. This
simple form has the advantage of preserving the self-similarity of the
collapse, which provides a powerful check on our numerically obtained 
solutions. We implement the ambipolar diffusion into the ZeusMP MHD 
code \citep{2006ApJS..165..188H} using the fully explicit method 
of \citet{1995ApJ...442..726M}. As 
is well known, for such explicit methods to be stable, the limiting 
timestep must be proportional to the grid size squared, which puts 
stringent constraints on spatial resolution. 

We solve the governing MHD equations including ambipolar diffusion 
in a spherical coordinate system ($r, \theta, \phi$) assuming
axisymmetry. The grid spacing is logarithmic in radius 
$r$ and constant in polar angle $\theta$, 
with a computational domain extending radially from $10^{14}$ to 
$2\times10^{17}$~cm and 0 to $\pi$ in angle. There are 120 grid points in 
the radial direction, and 60 angular points. The smallest grid size 
is $\sim 10^{13}$~cm, near the inner boundary. The standard 
outflow condition is imposed at the outer radial boundary. The inner 
boundary is a modified outflow boundary with the mass accreted across 
the boundary added to the central point mass and a torque free 
condition imposed on the magnetic field (i.e., $B_\phi=0$, see 
Paper I). We adopt the oft-used broken-power law equation of state 
that is isothermal below $\rho=10^{-13}$~g~cm$^{-3}$ and adiabatic 
with $\gamma=7/5$ above.


%
%

\section{Standard Model}

%
%
We first illustrate the effects of ambipolar diffusion on magnetic 
braking and disk formation using a model with a fiducial cosmic 
ray ionization rate $\zeta=10^{-17}$~s$^{-1}$ and a mass-to-flux 
ratio $\lambda=4$; the latter is close to the mean value inferred by 
\citet{2008ApJ...680..457T} in their Arecibo OH Zeeman observations 
of dark cloud cores. Other combinations of $\zeta$ and $\lambda$ 
are considered in the next section. For all models, we set the  
sound speed to $c_s=0.3$~km/s and the rotational speed to 
$v_0=c_s/2=0.15$~km/s, which corresponds to an angular speed 
of $3$~km~s$^{-1}$~pc$^{-1}$, typical of dense cores \citep{1993ApJ...406..528G}.

%
%

\begin{figure}
\epsscale{0.75}
\plotone{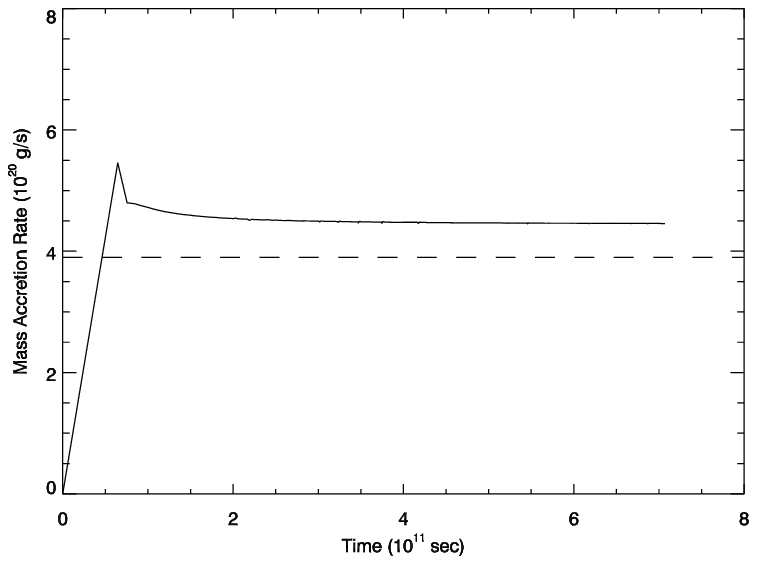}
\caption{The mass accretion rate of the standard model (solid line)
  compared to Shu's value for singular isothermal sphere (dashed). 
} 
\label{mdot}
\end{figure}

A quantity of primary importance to star formation is the mass 
accretion rate, ${\dot M}$. In Fig.~\ref{mdot}, we show ${\dot M}$
as a function of time. In our standard model, ${\dot M}$
converges to a constant value, after a short period of adjustment. 
The initial deviation from the
constant is due to the zero infall velocity at $t=0$ and the small
point mass that we put at the center to induce the collapse. The
accretion rate quickly approaches a constant value indicating the
solution has reached the expected self similar state at late
times. The converged value of ${\dot M}$ is close to that
of \citet{1977ApJ...214..488S} for a singular isothermal sphere. One 
might naively expect the rotation to retard the
collapse significantly, leading
to a reduced mass accretion rate onto the central object. Apparently,
the magnetic braking is strong enough that rotation is not a
significant barrier to mass accretion, even in the presence of
ambipolar diffusion. The accretion rate is comparable to that in the
ideal MHD case (Paper I), suggesting that the strength of braking is 
not significantly reduced by ambipolar diffusion. The accretion is 
highly episodic in the ideal MHD case, but is steady in the presence 
of ambipolar diffusion.

\begin{figure}
\epsscale{0.75}
\plotone{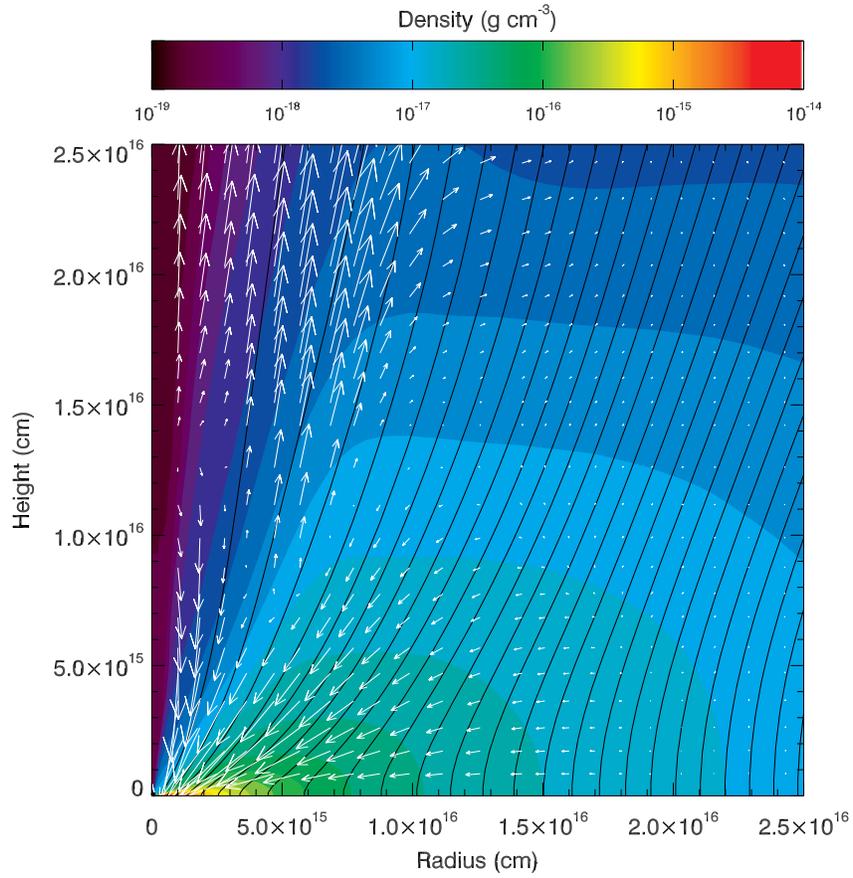}
\caption{Color map of density for the standard collapse solution, with
  velocity vectors showing simultaneous infall and outflow. The
  magnetic field lines (black contours) show strong pinching at small 
radii due to the collapse. The flattened high density region is a 
collapsing pseudodisk rather than a rotationally supported disk. 
} 
\label{DensMap}
\end{figure}

Fig.~\ref{DensMap} shows a snapshot of the standard model at a representative
time of $5.85\times10^{11}$~sec. As in the singular isothermal sphere,
the collapse occurs inside out, with the bulk of the envelope material beyond the
radius of $c_st=2\times10^{16}$~cm remaining nearly static. Inside this
radius, the matter accelerates towards the center. The collapsing flow
is deflected towards the equator by pinched field lines, forming a
pseudodisk \citep{1993ApJ...417..243G}. As the flow
collapses, it is expected to spin up due to conservation of angular
momentum. However, the rotation speed tends to decrease as the
collapsing material moves inward, especially in the equatorial
region, indicating significant loss of angular momentum due to 
magnetic braking.
%
%

To illustrate the braking more quantitatively, we plot the equatorial 
velocity profiles in Fig.~\ref{standard}. As one moves
from large radii to the center we observe four distinct regions in the
radial velocity profile (top panel). Outside of $r=2\times10^{16}$~cm the
matter remains nearly static. In the region $6\times10^{15}$~cm$ \lsim r \lsim
2\times10^{16}$~cm the matter accelerates toward the center as gravity
dominates the magnetic forces, since the mass-to-flux ratio of the
collapsing core is
significantly greater than unity. The collapsing material is
decelerated in the region $3\times10^{15}$~cm$ \lsim r \lsim
6\times10^{15}$~cm, where the outward magnetic forces exceed the
inward gravitational pull. This region does not exist in the
  ideal MHD limit; its presence is due entirely to ambipolar
  diffusion, which enables the magnetic field lines that would have
  been dragged into the central object in the ideal MHD limit to pile 
up outside the object  
\citep{1996ApJ...464..373L,1998ApJ...504..257C}. 
 In this decelerating
matter, the ion-neutral drift velocity becomes large because of a
large Lorentz force. At $r\sim 3\times10^{15}$~cm the neutral matter
begins accelerating towards the center again. However, the ions
remain nearly static at $r\sim1.5\times10^{15}$~cm, indicating that 
the magnetic
field lines barely move in that region. 
The deceleration region 
has the same physical origin as the hydromagnetic accretion shock induced 
by ambipolar diffusion \citep{1996ApJ...464..373L,1998ApJ...504..257C}. 
However, since the gas outside of the decelerating region 
is collapsing subsonically, a hydromagnetic 
accretion front rather than a shock is formed.
At $r \lsim 10^{15}$~cm the
ions (and magnetic field) begin reaccelerating towards the center. The
rapid collapse of both ions and neutrals at the smallest distances
from the star clearly suggest the absence of a rotationally supported 
disk.

From the lower panel of
Fig.~\ref{standard}, it is clear that the rotation 
speed $v_\phi$ does not increase with
decreasing radius, which is opposite of what is expected based on
angular momentum conservation. For example, in the region of
accelerating collapse ($6\times10^{15}$~cm$ \lsim r \lsim
2\times10^{16}$~cm), $v_\phi$ remains constant while the radius
decreases by a factor of $\sim 3$. This indicates significant loss 
of angular momentum. This loss can be understood by examining the 
ratio of toroidal to poloidal magnetic fields.
We find that the magnetic field lines get more twisted as the
collapse accelerates inwards. When deceleration occurs, compression
increases the poloidal and toroidal field strength, which leads 
to a stronger braking that is responsible for the steep drop in 
rotational speed in the deceleration region (the shaded region 
in Fig.~\ref{standard}). By the
time the fluid reaccelerates towards the center, there is little
angular momentum left in the gas to maintain a significant twist in the 
magnetic field, resulting
in near radial infall to the center. The supersonic infall velocity
and small rotation speed at small radii leaves no doubt that
rotationally supported disk formation is suppressed in our standard 
model.

\section{Effects of Ionization Rate and Magnetic Field Strength}

%
%
Our collapse solution depends on the cosmic ray ionization rate, which 
is somewhat uncertain. For dense cores, the typical values lie in a
range between $(1-5)\times 10^{-17}$~s$^{-1}$ \citep[e.g.,][]{2007ApJ...664..956M}, 
although a wider spread is also possible \citep{1998ApJ...499..234C}. 
To examine the sensitivity of our results to the ionization rate, we vary $\zeta$ 
by a factor of $10$ in either direction. 

The higher ionization rate case has a larger ionization fraction by a
factor of $\sqrt{10}$, which leads to stronger coupling between the
matter and magnetic field. Its collapse is qualitatively similar to 
the standard model, with a few
quantitative differences. The stronger coupling allows the magnetic
fields to be brought closer to the central object, leaving a larger
magnetic flux in a smaller region. As a result, the deceleration
region is closer to the central object (see dashed lines of the 
top panel of Fig.~\ref{variations}). The stronger field and stronger 
coupling keeps the ion radial speed low
for a more extended region during the reacceleration. The stronger
field also results in enhanced braking which removes essentially 
all the angular momentum at small radii (see dashed lines of the second 
panel, noting that the rotation speeds for ions and neutrals are practically
indistinguishable). 
Increasing the ionization rate further suppresses rotationally
supported disk formation.

The lower ionization rate yields an ionization fraction smaller by a 
factor of $\sqrt{10}$, which leads to weaker coupling between the
matter and magnetic field. The ions start to deccelerate earlier, 
due to the pile up of magnetic flux at larger radii. However, the 
neutrals slip
through the ions without being appreciably deccelerated due to weak
coupling, resulting in faster diffusion. This results in a faster 
collapse of the neutrals and a
slower collapse of the ions compared to the standard model 
(see the dot-dashed lines in top panel). The increased
diffusion requires a smaller ambipolar diffusion timestep, which 
increases computational time\footnote{Although the two models of 
$\zeta=10^{-18}$~s$^{-1}$ in Fig.~\ref{variations} were evolved for a
shorter time ($2.95\times10^{11}$~s compared to 
$5.85\times10^{11}$~s for other, better coupled models), they 
still converged to 
a self similar solution. Their velocity profiles  
have been self similarly scaled for comparison.}. The weak 
coupling and reduced magnetic field strength decrease
the strength of magnetic braking, allowing the neutrals to spin up 
as they collapse. However, there is still enough braking to prevent 
a rotationally supported disk from forming inside our computational
domain, as evidenced by the rapid radial collapse at small radii. 
We conclude that for moderately strongly magnetized core of
$\lambda=4$, disk formation is suppressed for realistic values of 
the cosmic ray ionization rate.

The mass-to-flux ratio of dense cores is uncertain. Our standard
model is based on the mean mass-to-flux ratio inferred by 
\citet{2008ApJ...680..457T}, but many cores have only lower limits 
on the mass-to-flux ratio. We investigate the collapse model in more 
weakly magnetized cores with $\lambda=13.3$ (same as the standard 
ideal MHD case of Paper I), which corresponds to 
a field strength of $7.35~\mu G$ on the scale of the typical core 
radius of 0.05~pc. We take this case as a lower 
limit of realistic magnetic field strengths based on the median 
field strength of cold neutral HI 
structures of $\sim 6~\mu G$ inferred by \citet{2005ApJ...624..773H}.

%
%
We again consider three values for the ionization rate $\zeta=
10^{-16}$, $10^{-17}$, and $10^{-18}$~s$^{-1}$, concentrating 
first on the case with the fiducial value of $10^{-17}$~s$^{-1}$. 
In contrast to the standard model with the same $\zeta$, 
there are two equatorial decceleration regions in this weaker 
field case instead of one (see solid lines in the third  
panel of Fig.~\ref{variations}). The accelerating infall is first 
decelerated near $r \sim 10^{16}$~cm.
This deceleration is {\it not} a new feature caused by 
ambipolar diffusion; it was already present in our previous ideal 
MHD calculation: it is a ``magnetic barrier'' caused by the 
bunching of magnetic field lines at the interface between the 
collapsing envelope and an expanding, magnetic braking-driven 
bubble (see figure~3 of Paper I). Inside the barrier, the
material recollapses inwards, at an increasingly high speed, 
until a radius of $\sim 10^{15}$~cm, where it is decelerated 
for a second time. The second deceleration {\it is} caused by
ambipolar diffusion, which allows magnetic flux to pile up 
at small radii, instead of being dragged into the center. 
As in the standard model, the magnetic pressure in the 
deceleration region becomes comparable to the ram pressure 
of the collapsing material. However, in this model the deceleration 
region occurs at a smaller radius, allowing the infalling gas 
to collapse supersonically, causing a hydromagnetic shock instead
 of a front.
The slowdown of infall compresses the 
collapsing material, leading to a stronger poloidal and 
toroidal magnetic field that enhance the rate of magnetic 
braking. The enhancement is responsible for the dip on the  
rotational speed profile (see solid lines in
bottom panel). Interior to the 
deceleration region, the collapsing flow accelerates for 
a third time, spinning up as it collapses. Nevertheless, 
there is insufficient angular momentum left in the 
material for a rotationally supported disk to form before 
the inner boundary is reached.

The dynamics of the more ionized, $\zeta=10^{-16}$~s$^{-1}$ 
case is similar to that of the $\zeta=10^{-17}$~s$^{-1}$ 
case down to the magnetic diffusion-induced second deceleration
near $\sim 10^{15}$~cm. Near this region and interior to it, 
the infall speed is also similar, but the rotational speed is
drastically different (see dashed lines in the bottom two 
panels). The field compression in the deceleration 
region increases the rate of magnetic braking to such a degree
as to cause a {\it counter-rotation} in the better coupled 
case. The better field-matter coupling and stronger magnetic
field lead to efficient removal of angular momentum at small 
radii, suppressing disk formation completely. 

Only in the most weakly ionized case of $\zeta=10^{-18}$~s$^{-1}$
does a rotationally support disk form. The disk is most obvious 
in the radial velocity curve (dot-dashed lines in the third panel). 
As in the more strongly coupled cases, there is accelerating collapse 
at large radii. The collapse is only weakly slowed by the magnetic 
barrier, because the neutrals are weakly coupled to the field. The 
ions and neutrals are reaccelerated, then decelerated for a second 
time. Unlike the more strongly coupled cases, this second 
deceleration is 
{\it not} due to the ambipolar diffusion-enabled pile up of magnetic flux, 
but is caused by centrifugal force from rapid rotation (see 
dot-dashed lines in the bottom 
panel). The combination of a weak magnetic
field and weak matter-field coupling renders the braking too 
inefficient to remove enough angular momentum to suppress disk 
formation in this case of extreme parameters. 

\section{Discussion and Conclusion}

\begin{figure}
\epsscale{0.75}
\plotone{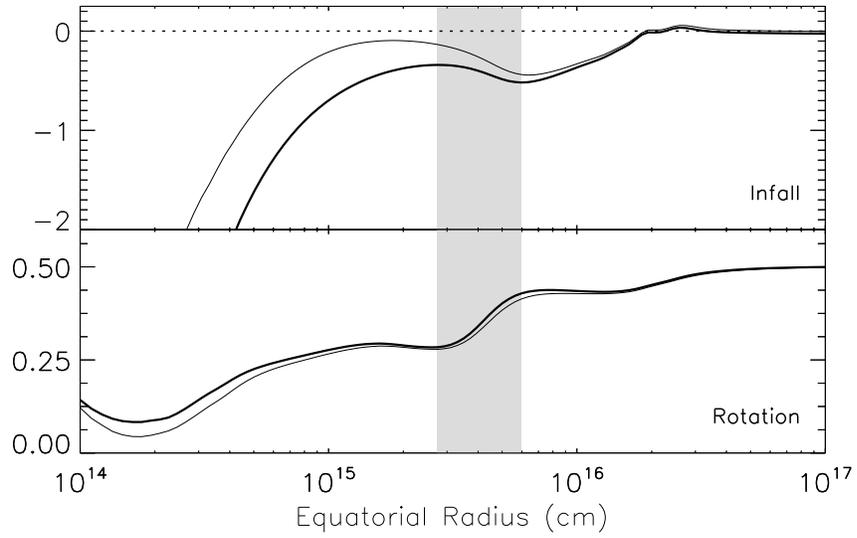}
\caption{Radial (top panel) and rotational (bottom) velocities in
  units of the sound speed of the 
neutrals (thick) and ions (thin) on the equator for the standard 
model. Note the decrease of the rotation velocity with decreasing 
radius, indicating strong magnetic braking, especially in the 
deceleration region (shaded). 
} 
\label{standard}
\end{figure}

\begin{figure}
\epsscale{0.65}
\plotone{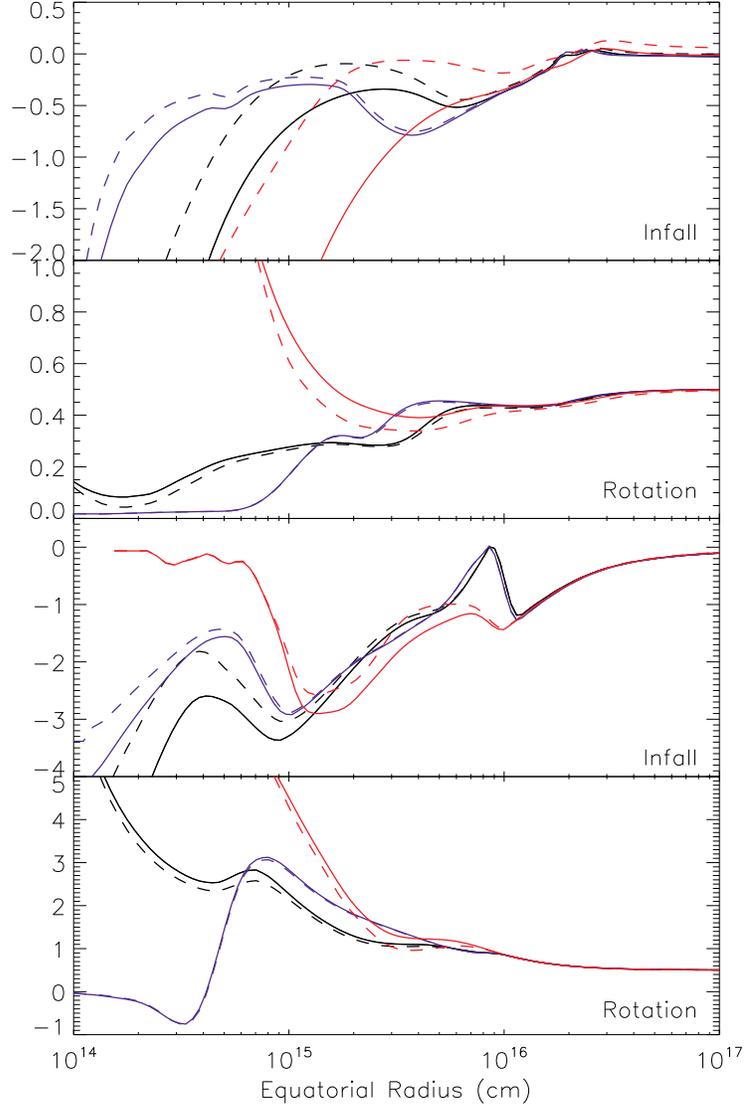}
\caption{Velocity profiles of different collapse models. The 
top panel shows the neutral (solid) and ion 
(dashed) radial velocities in unit of the sound speed for three 
models with the same core mass-to-flux ratio $\lambda =4$ but different 
cosmic ray ionization rates $\zeta=10^{-16}$ (blue), 
$10^{-17}$ (black), and
$10^{-18}$~s$^{-1}$ (red).
The rotation speeds
of these models are shown in the second panel. The third and bottom 
panels show, respectively, the radial and rotation velocities of 
three models with a weaker magnetic field of $\lambda=13.3$ but 
different ionization rates $\zeta=10^{-16}$ (blue), 
$10^{-17}$ (black), and
$10^{-18}$~s$^{-1}$ (red). Note that the
rotationally supported disk forms only in the model of 
the lowest ionization rate {\it and} weaker magnetic field.} 
\label{variations}
\end{figure}

We have studied the collapse of rotating molecular cloud cores 
magnetized to moderate degrees, concentrating on a dimensionless 
mass-to-flux ratio $\lambda=4$, as suggested by recent Zeeman 
observations \citep{2008ApJ...680..457T}. In the ideal MHD limit, 
we have previously shown 
that a slight twist of the field lines in such cores is 
sufficient to remove essentially all of the angular momentum of 
the collapsing matter and suppress the formation of rotationally
supported disks during the protostellar accretion phase completely 
\citep[Paper I ; see also][]{2006ApJ...647..374G,2008A&A...477....9H}. 
One may expect ambipolar diffusion to 
reduce the efficiency of magnetic braking and potentially enable 
the disks to form, since the twisting of field lines by the 
rotation of neutrals will be reduced by diffusion. However, for 
realistic cosmic ray ionization rates, the small amount of field 
twisting needed for angular momentum removal drives a drift speed 
between the ions (and the field lines tied to them) and neutrals 
in the azimuthal direction that is much smaller than the neutral 
rotation speed (see the lower panel of Fig.~\ref{standard}). 
By itself, the azimuthal slippage 
of field lines relative to neutrals does not reduce the 
strength of the toroidal magnetic field (and the braking rate) 
significantly. The ambipolar diffusion is expected to modify the 
poloidal magnetic field as well, especially in regions where the 
magnetic force is large. The modification is most evident at 
small radii, where magnetic diffusion has allowed the magnetic 
flux that would have been dragged onto the central object to 
occupy an extended region \citep{1996ApJ...464..373L,1998ApJ...504..257C,2007ApJ...660..388T}. 
The increased poloidal field  
strength tends to make the magnetic braking more efficient  
\citep{2002ApJ...580..987K}. Given these competing 
effects in opposite directions, it not at all obvious whether 
ambipolar diffusion would increase or decrease the overall 
efficiency of magnetic braking. Our detailed calculations have 
shown that, for realistic rates of cosmic ray ionization, the 
overall efficiency is increased, because the azimuthal field 
slippage is small and the poloidal field trapped at small radii
is strong. The efficiency is reduced only for unrealistically 
low ionization rates, which speed up the azimuthal field 
slippage and lower the poloidal field strength by enabling the 
trapped flux to escape to large distances. 

There is, however, ample evidence for large, rotationally supported 
disks on $10^2$~AU scale or more around young stars, at least in 
Class I and II, and possibly in Class 0, sources. This begs the
question: what is the origin of these disks?

The most obvious possibility is that additional non-ideal MHD effects
may weaken the magnetic braking further. Ohmic dissipation is 
expected to be important at high densities \citep{2006ApJ...647..382S}. \citet{2002ApJ...573..199N} 
estimates that it dominates other processes at densities above 
$\sim 10^{12}$~cm$^{-3}$ \citep[see also][]{2001ApJ...550..314D}. 
Since the density in our standard model never exceeds this
value, we believe that Ohmic dissipation would not significantly 
weaken the magnetic braking inside our computation domain (from 
$10^{14}$~cm$ < r < 2\times10^{17}$~cm). However, Ohmic dissipation 
may still enable a rotationally supported disk to form at a smaller 
radius where the density is higher, as demonstrated explicitly by 
\citet{2008ApJ...676.1088M} using a resistive MHD core. They found a 
disk of $\sim10^{12}$~cm in radius shortly ($\sim 11$~days) after 
the formation of the so-called ``second core''; the disk was not 
present in the ideal MHD counterpart. Such Ohmic dissipation-enabled 
disks can form inside our inner boundary (of $10^{14}$~cm). Since
the material crossing the inner boundary of our simulation is 
typically braked to
a rotational speed well below the local Keplerian speed, any disk 
that may form later should be relatively small in the absence of 
angular momentum redistribution.

Besides Ohmic dissipation, another non-ideal mechanism of magnetic flux
diffusion is the Hall Effect. When dust grains abundances are low, the
Hall Effect becomes important when collisions with neutrals decouple 
the ions (but not electrons) from the magnetic field. This occurs 
when the ion Hall parameter $\beta_i$ is less than unity \citep[e.g.,][]{1984FCPh....9..139N}. 
We have checked that $\beta_i>1$ everywhere inside our 
computational domain for the standard model. We do not expect the 
Hall Effect to change our results significantly, although this 
may be changed by the inclusion of dust grains. The exact effect 
depends on the grain size distribution \citep{1999MNRAS.303..239W}, which 
is uncertain. Further study is required to determine the importance of the Hall Effect. 
If the additional non-ideal MHD
effects do not allow the observed large scale disks to form, other
means of weakening the magnetic braking must be sought.

One possibility is protostellar outflow. Protostellar outflows are
observed at all stages of star formation. They are thought to clear
away the envelope material as the source ages \citep{2006ApJ...646.1070A}. As
the envelope is the anchor for the magnetic brake in our models,
removing it may reduce the strength of braking sufficiently to allow
large scale rotationally supported disks to form. This is particularly
true for the more evolved class II and perhaps class I sources, where 
little envelope is left to brake the disk. It may not work for class 
0 sources where the outflows tend to be confined to the polar regions 
(e.g., HH 211, \citealp{1999A&A...343..571G}). If this is the
case, we would not expect these deeply embedded objects to harbor
large scale, rotationally supported disks. They could still have 
relatively small disks enabled by, for example, Ohmic dissipation. 
It is difficult to determine the exact amount of rotational support 
in the equatorial region from current observations of Class 0 sources. 
The situation will improve when ALMA comes
online.

\acknowledgements
We thank Mordicai Mac Low and Jeff Oishi for help in the implementation 
of ambipolar diffusion in ZEUS MP. This work is supported in part by NASA 
(NNG05GJ49G) and NSF (AST-0307368) grants.

\clearpage

\begin{thebibliography}{}
\bibitem[Adams \& Shu(2007)]{2007ApJ...671..497A} Adams, F.~C., \& Shu, F.~H.\ 2007, \apj, 671, 497 
\bibitem[Arce \& Sargent(2006)]{2006ApJ...646.1070A} Arce, H.~G., \& Sargent, A.~I.\ 2006, \apj, 646, 1070 
\bibitem[Allen et al.(2003)Allen, Li, \& Shu]{2003ApJ...599..363A} Allen, A., Li, Z.-Y., \& Shu, F.~H.\ 2003, \apj, 599, 363 
\bibitem[Banerjee \& Pudritz(2006)]{2006ApJ...641..949B} Banerjee, R., \& Pudritz, R.~E.\ 2006, \apj, 641, 949 
\bibitem[Basu \& Mouschovias(1994)]{1994ApJ...432..720B} Basu, S., \& Mouschovias, T.~C.\ 1994, \apj, 432, 720 
\bibitem[Bodenheimer(1995)]{1995ARA&A..33..199B} Bodenheimer, P.\ 1995, \araa, 33, 199 
\bibitem[Boss(1998)]{1998ASPC..148..314B} Boss, A.~P.\ 1998, Origins, 148, 314 
\bibitem[Caselli et al.(1998)]{1998ApJ...499..234C} Caselli, P., Walmsley, C.~M., Terzieva, R., \& Herbst, E.\ 1998, \apj, 499, 234 
\bibitem[Ciolek \& K{\"o}nigl(1998)]{1998ApJ...504..257C} Ciolek, G.~E., \& K{\"o}nigl, A.\ 1998, \apj, 504, 257 
\bibitem[Desch \& Mouschovias(2001)]{2001ApJ...550..314D} Desch, S.~J., \& Mouschovias, T.~C.\ 2001, \apj, 550, 314 
\bibitem[Dib et al.(2007)]{2007ApJ...661..262D} Dib, S., Kim, J., V{\'a}zquez-Semadeni, E., Burkert, A., \& Shadmehri, M.\ 2007, \apj, 661, 262 
\bibitem[Elmegreen(1979)]{1979ApJ...232..729E} Elmegreen, B.~G.\ 1979, \apj, 232, 729 
\bibitem[Galli et al.(2006)]{2006ApJ...647..374G} Galli, D., Lizano, S., Shu, F.~H., \& Allen, A.\ 2006, \apj, 647, 374 
\bibitem[Galli \& Shu(1993)]{1993ApJ...417..243G} Galli, D., \& Shu, F.~H.\ 1993, \apj, 417, 243 
\bibitem[Girart et al.(2006)Girart, Rao, \& Marrone]{2006Sci...313..812G} Girart, J.~M., Rao, R., \& Marrone, D.~P.\ 2006, Science, 313, 812 
\bibitem[Goodman et al.(1993)]{1993ApJ...406..528G} Goodman, A.~A., Benson, P.~J., Fuller, G.~A., \& Myers, P.~C.\ 1993, \apj, 406, 528 
\bibitem[Gueth \& Guilloteau(1999)]{1999A&A...343..571G} Gueth, F., \& Guilloteau, S.\ 1999, \aap, 343, 571
\bibitem[Hayes et al.(2006)]{2006ApJS..165..188H} Hayes, J.~C., Norman, M.~L., Fiedler, R.~A., Bordner, J.~O., Li, P.~S., Clark, S.~E., ud-Doula, A., \& Mac Low, M.-M.\ 2006, \apjs, 165, 188 
\bibitem[Heiles \& Troland(2005)]{2005ApJ...624..773H} Heiles, C., \& Troland, T.~H.\ 2005, \apj, 624, 773 
\bibitem[Hennebelle \& Fromang(2008)]{2008A&A...477....9H} Hennebelle, P., \& Fromang, S.\ 2008, \aap, 477, 9 
\bibitem[Krasnopolsky \& K{\"o}nigl(2002)]{2002ApJ...580..987K} Krasnopolsky, R., K{\"o}nigl, A.\ 2002, \apj, 580, 987 
\bibitem[Li \& McKee(1996)]{1996ApJ...464..373L} Li, Z.-Y., \& McKee, C.~F.\ 1996, \apj, 464, 373 
\bibitem[Lizano \& Shu(1989)]{1989ApJ...342..834L} Lizano, S., \& Shu, F.~H.\ 1989, \apj, 342, 834 
\bibitem[Mac Low et al.(1995)]{1995ApJ...442..726M} Mac Low, M.-M., Norman, M.~L., Konigl, A., \& Wardle, M.\ 1995, \apj, 442, 726 
\bibitem[Machida et al.(2008)]{2008ApJ...676.1088M} Machida, M.~N., Inutsuka, S.-i., \& Matsumoto, T.\ 2008, \apj, 676, 1088 
\bibitem[Machida et al.(2006)]{2006ApJ...647L.151M} Machida, M.~N., Inutsuka, S.-i., \& Matsumoto, T.\ 2006, \apjl, 647, L151 
\bibitem[Maret \& Bergin(2007)]{2007ApJ...664..956M} Maret, S., \& Bergin, E.~A.\ 2007, \apj, 664, 956 
\bibitem[Mellon \& Li(2008)]{2008ApJ...681.1356M} Mellon, R.~R., \& Li, Z.-Y.\ 2008, \apj, 681, 1356
\bibitem[Mouschovias \& Ciolek(1999)]{1999osps.conf..305M} Mouschovias, T.~C., \& Ciolek, G.~E.\ 1999,in NATO ASIC Proc.~540, The Origin of Stars and Planetary Systems, ed Lada, C.~J. and Kylafis, N.~D. (Norwell, MA: Kluwer Acad. Pub.), 305 
\bibitem[Nakamura \& Li(2008)]{2008arXiv0804.4201N} Nakamura, F., \& Li, Z.-Y.\ 2008, in press 
\bibitem[Nakano(1984)]{1984FCPh....9..139N} Nakano, T.\ 1984, Fundamentals of Cosmic Physics, 9, 139 
\bibitem[Nakano \& Nakamura(1978)]{1978PASJ...30..671N} Nakano, T., \& Nakamura, T.\ 1978, \pasj, 30, 671 
\bibitem[Nakano et al.(2002)Nakano, Nishi, \& Umebayashi]{2002ApJ...573..199N} Nakano, T., Nishi, R., \& Umebayashi, T.\ 2002, \apj, 573, 199 
\bibitem[Price \& Bate(2007)]{2007Ap&SS.311...75P} Price, D.~J., \& Bate, M.~R.\ 2007, \apss, 311, 75 
\bibitem[Shu(1977)]{1977ApJ...214..488S} Shu, F.~H.\ 1977, \apj, 214, 488 
\bibitem[Shu et al.(2006)]{2006ApJ...647..382S} Shu, F.~H., Galli, D., Lizano, S., \& Cai, M.\ 2006, \apj, 647, 382 
\bibitem[Tassis \& Mouschovias(2007)]{2007ApJ...660..388T} Tassis, K., \& Mouschovias, T.~C.\ 2007, \apj, 660, 388 
\bibitem[Terebey et al.(1984)Terebey, Shu, \& Cassen]{1984ApJ...286..529T} Terebey, S., Shu, F.~H., \& Cassen, P.\ 1984, \apj, 286, 529 
\bibitem[Tilley \& Pudritz(2005)]{2005prpl.conf.8473T} Tilley, D.~A., \& Pudritz, R.~E.\ 2005, Protostars and Planets V, 8473 
\bibitem[Tomisaka(1998)]{1998ApJ...502L.163T} Tomisaka, K.\ 1998, \apjl, 502, L163 
\bibitem[Troland \& Crutcher(2008)]{2008ApJ...680..457T} Troland, T.~H., \& Crutcher, R.~M.\ 2008, \apj, 680, 457 
\bibitem[Ward-Thompson et al.(2000)]{2000ApJ...537L.135W} Ward-Thompson, D., Kirk, J.~M., Crutcher, R.~M., Greaves, J.~S., Holland, W.~S., \& Andr{\'e}, P.\ 2000, \apjl, 537, L135 
\bibitem[Wardle \& Ng(1999)]{1999MNRAS.303..239W} Wardle, M., \& Ng, C.\ 1999, \mnras, 303, 239 





\end{thebibliography}
\end{document}